\documentclass[11pt]{article}
\usepackage[scale=.72,centering]{geometry}
\usepackage{amssymb,amsmath}
\usepackage{latexsym}
\usepackage[round]{natbib}
\usepackage{fleqn,latexsym}
\usepackage{smallsec,nustyle2}
\usepackage{endnotes}
\usepackage{multirow}
\usepackage[dvipdf]{color}

\usepackage{graphicx}
\usepackage{color}
\definecolor{Blue}{rgb}{0.2,0.3,0.9}
\definecolor{Green}{rgb}{0.0,0.7,0.0}

\newlength{\wrapsep}
\newlength{\saveintextsep}
\definecolor{myGreen}{rgb}{0.0,0.0,0.7}
\definecolor{myBlack}{rgb}{0.0,0.0,0.0}
\definecolor{myRed}{rgb}{0.5,0.0,0.00}

\newtheorem{prop}[equation]{Proposition: }
\newtheorem{opqu}[equation]{Open Question: }
\newtheorem{cor}[equation]{Corollary: }

\newtheorem{defn}[equation]{Definition: }

\newtheorem{dispar}[equation]{}

\newcommand{\BB}{\ensuremath{\square}}
\newcommand{\BD}{\ensuremath{\lozenge}}
\newcommand{\SIGN}{\ensuremath{\mathbb{L}}}
\newcommand{\UIND}{\ensuremath{\mathbb{U}}}

\newcommand{\ASIGN}{\ensuremath{(\SIGN,\UIND)}}
\newcommand{\LANG}{\ensuremath{\mathcal{L}}}

\newcommand{\FANG}{\ensuremath{\LANG\ASIGN}}
\newcommand{\WORLDS}{\ensuremath{\mathbb{W}}}

\newcommand{\BMOD}{\ensuremath{\langle D,\WORLDS,t,u,s\rangle}}
\newcommand{\PARMOD}{\ensuremath{\langle D,\WORLDS,t,u,s_n\rangle}}
\newcommand{\BNOD}{\ensuremath{\langle D,\WORLDS,t\rangle}}

\newcommand{\UU}{\ensuremath{u}} 
\newcommand{\WF}[2]{\ensuremath{#1[\,#2\,]}}
\newcommand{\WTF}[3]{\ensuremath{#1[\,#2,#3\,]}}
\newcommand{\MO}[1]{\mbox{$\mathcal{#1}$}}
\newcommand{\TAUT}{\ensuremath{\top}}
\newcommand{\CONT}{\ensuremath{\bot}}
\newcommand{\QED}{\ensuremath{\hfill \Box}}
\newcommand{\MID}{\mbox{$\;\mid\;$}}
\newcommand{\NATN}{\ensuremath{\boldmath{N}}}
\newcommand{\ASSI}{\ensuremath{d}}
\newcommand{\NICEDEF}{\ensuremath{\overset{\normalfont{\text{def}}}{=}}}
\newcommand{\PRO}{\ensuremath{\unrhd}}
\newcommand{\BOOD}{\ensuremath{\langle D,\WORLDS,t,\PRO,s\rangle}}
\newcommand{\GRO}{\ensuremath{\boldmath{r}}}
\newcommand{\GOOD}{\ensuremath{\langle D,\WORLDS,t,\GRO\rangle}}
\newcommand{\trans}[1]{\mbox{${#1}^{\dagger}$}}
\newcommand{\RELF}[1]{\mbox{$\mathcal{F}_{\mathcal{#1}}$}}
\newcommand{\PFAM}{\ensuremath{\mathfrak P}}

\newcommand{\TRAL}{transitivity}
\newcommand{\TRA}{transitive}

\newcommand{\INVU}{utility-invariant}

\newcommand{\INVUL}{utility-invariance}

\renewcommand{\baselinestretch}{1.02}
\title{Quantified preference logic\thanks{
Contact: osherson@princeton.edu, weinstein@cis.upenn.edu}}

\author{%
Daniel Osherson \\ Princeton University \and
Scott Weinstein \\ University of Pennsylvania}
\begin{document}
\maketitle
\begin{abstract}\noindent
The logic of reason-based preference advanced in
\cite{PBOR} is extended to quantifiers. Basic properties of the
new system are discussed.
\end{abstract}

\section{Introduction}

In \cite{PBOR} we proposed a logic with binary modal connective $\succeq_X$
that can be read informally as follows.
\begin{quote}
Let formulas $\varphi$ and $\psi$ be given along with a collection $X$ of utility
scales that measure value along various dimensions. Then
$\varphi \succeq_X \psi$ iff the situation that comes to mind
when envisioning $\varphi$ being true is weakly preferred to the situation
that comes to mind when envisioning $\psi$ being true, according
to the criteria indexed by $X$.
\end{quote}

\noindent
The utility scales can be conceived as ``reasons-for-preference.''
For example, $p$ might be dinner tonight with Mitt Romney, and $q$ dinner
tonight with Barack Obama. Then if $X$ consists of the two reasons ``enjoy
lively conversation,'' and ``influence policy,'' it might be true of you that
$q \succeq_X (p \wedge q)$. Note that the two reasons cut in different
directions: dinner with both would be more lively but offer less chance to
influence policy. Affirming the formula thus results from aggregating the
two considerations. Of course, you might instead have to evaluate just
$p \succeq_X \TAUT$, that is, whether dinner with Mitt is weakly better than
your present situation (represented by the tautology \TAUT).

Other examples are discussed in
\cite{PBOR}, and a formal semantics for such sentences is presented and
analyzed. The system can be seen as a propositional calculus extended with
modal binary connectives. Our present purpose is to show how the propositional
part can be replaced with predicate calculus.
It is left to \cite{PBOR} to indicate the
previous work that inspired ours, notably, \cite{Liu08,List}.
To get started, we specify the languages under consideration, then turn to
semantics.

\section{Language}

\subsection{Signatures}

A quantified language is built from its ``signature.''

\begin{defn}\label{sem2}
By a \textit{signature} is meant a pair \ASIGN\ where
\begin{enumerate}
\item\label{sem2a} \SIGN\ is a
collection of
predicates and function symbols of various arities.
\item\label{sem2b} \UIND\ is a nonempty collection
of nonempty subsets of
natural numbers ($0, 1 \dots$).
\end{enumerate}
\end{defn}
The numbers appearing in $X\in\UIND$ represent specific
reasons for preference such as the desire for
lively conversation in our example. A set $X$ of reasons
influences preference through aggregation of its members.
If $\bigcup\UIND\in\UIND$ then preference according to
$\bigcup\UIND$ amounts to preference \textit{tout court}.

\subsection{Formulas}

\noindent
We specify the
language \FANG\ parameterized by the signature \ASIGN.  Formulas
are built from the following symbols.

\vspace*{-2mm}

\begin{enumerate}\setlength{\itemsep}{-1mm}
\item the members of \SIGN\ along with the identity sign $=$
\item for each $X\in\UIND$, the binary connective $\succeq_X$
\item the binary connective $\wedge$ and the unary connective $\neg$
\item the quantifier $\exists$
\item the two parentheses, $($, $)$
\item a denumerable collection $v_0, v_1 \dots$ of individual variables
(denoted below by $x$, $y$, $z$).
\end{enumerate}

\vspace*{-2mm}

\noindent
The set of \textit{terms} is constructed from functions and variables
as usual. The set \FANG\ of \textit{formulas}
is likewise built in the usual way except that we add the clause:
\begin{quote}
Given $\varphi,\psi\in\FANG$ and $X\in\UIND$, $\varphi\succeq_X\psi$ also belongs
to \FANG.
\end{quote}

\noindent
Moreover, we rely on the following abbreviations.
\def\ABB{\text{for}}
\begin{eqnarray*}
\forall x\varphi & \ABB & \neg\exists x\neg\varphi\\
(\varphi\vee\psi) & \ABB & \neg(\neg\varphi \wedge\neg\psi)\\
(\varphi\rightarrow\psi) & \ABB & (\neg\varphi \vee\psi)\\
(\varphi\leftrightarrow\psi) & \ABB & ((\varphi
\rightarrow\psi)\wedge (\psi \rightarrow\varphi))\\
(\varphi\succeq_{1 \dots k}\psi) & \ABB & (\varphi\succeq_{\{1
\dots k\}}\psi)\\
(\varphi\succ_X\psi) & \ABB & (\varphi\succeq_X\psi)\wedge
\neg(\psi\succeq_X\varphi)\\
(\varphi\approx_X\psi) & \ABB & (\varphi\succeq_X\psi)\wedge
(\psi\succeq_X\varphi)\\
(\varphi\preceq_X\psi) & \ABB & (\psi\succeq_X\varphi)\\
(\varphi\prec_X\psi) & \ABB & (\psi\succ_X\varphi)\\
\TAUT & \ABB & \forall x(x=x)\\
\CONT & \ABB & \neg\TAUT\\
\end{eqnarray*}

\subsection{Examples}

The following formulas serve as illustration.

\begin{dispar}\label{opti1}
\begin{enumerate}
\item\label{opti1b} $\exists x(Px \succ_X \forall yPy)$
\item\label{opti1a} $\exists xPx \succ_X \forall yPy$
\end{enumerate}
\end{dispar}
\noindent
In the domain of people, \ref{opti1}\ref{opti1b} affirms that
there is someone for whom satisfying $P$ is preferable to
everyone satisfying it. This might well be true. For example,
from my perspective, it's better that I discover a metric ton
of gold than that everyone does (where the reasons encoded
in $X$ are basely materialistic). In contrast, \ref{opti1}\ref{opti1a}
entails that someone getting the gold is better than everyone getting it,
which might be false if it doesn't strike me as plausible that I'm
the lucky person.

The next example exhibits modal embedding. Consider the domain
of potential automobile purchases. Let $P$ refer to the purchase of
an efficient
gasoline car, and $Q$ to the purchase of a fully electric vehicle.
Finally, let the utility indices $1,2$ measure, respectively,
ecological value and financial interest. Then
\[
\forall x\,(\,Px\ \prec_1\ \exists y\,(Qy\ \succ_2\ Px\,)\,)
\]
says that buying an efficient gasoline car is not as
ecologically useful as there being an electric car whose purchase
is financially more attractive than buying the gas vehicle. The
formula is true if electricity is ecologically superior
to gas but consumer choice is based on narrow economic interest.
We now turn from intuitive meaning to formal semantics.

\section{Semantics}\label{semSec}

Recall that a signature \ASIGN\ consists of vocabulary
(\SIGN) and sets of utility indices (\UIND).
\subsection{Models}

\begin{defn}\label{sem1}
Let a signature \ASIGN\ be given.
By a \textit{model} for the signature is meant
a quintuple $\MO{M} = \BMOD$ where:
\begin{enumerate}
\item\label{sem1a} $D$ is a nonempty set, the \textit{domain} of \MO{M}.
\item\label{sem1b} \WORLDS\ is a nonempty set of points, the \textit{worlds} of \MO{M}.
\item\label{sem1c} $t$ maps $\WORLDS\times\SIGN$ to the appropriate set-theoretic objects
over $D$. (For example, if $Q\in\SIGN$ is a binary relation symbol then
$t(w,Q)$ is a subset of $D\times D$.) Identity is assigned to $=$.
\item\label{sem1e} $u$ is a function from $\UIND\times \WORLDS$ to the real numbers.
For $X,\{i\}\in\UIND$ we write $u_X(w)$ in place of $u(X,w)$
and $u_i(w)$ in place of $u(\{i\},w)$.
\item\label{sem1d} $s$ is a function from $\WORLDS\times \{A\subseteq \WORLDS\MID\emptyset\neq A\}$
such that for all $w\in \WORLDS$ and $\emptyset\neq A\subseteq W$, $s(w,A)\in A$.
\end{enumerate}
\end{defn}
\noindent
Thus, \WORLDS\ corresponds to a set of potential situations; via $t$, each gives extensions
in $D$ to the vocabulary in \SIGN. The function $u_X$ measures the utility of worlds
according to the considerations encoded in $X\in\UIND$. Finally, given a world $w_0$ and
a set $A$ of worlds, $s$ selects a ``cognitively salient'' member
of $A$, where salience may depend on the vantage point $w_0$.

\subsection{Propositions}

Subsets of worlds are called \textit{propositions}. In the context of a given model,
our semantic definition assigns a proposition (subset of \WORLDS) to each formula.

Fix a signature \ASIGN, and
let a model $\MO{M} = \BMOD$ be given.
By an \textit{assignment (for \MO{M})} is meant a map of the individual variables
of \FANG\ into $D$. Given a variable $x$ and assignment \ASSI, an $x$ \textit{variant} of
\ASSI\ is any assignment that differs from \ASSI\ at most in the member of $D$
assigned to $x$. Assignments are extended to terms of \FANG\ in the usual way.

\begin{defn}\label{meaningDef}
Let a model $\MO{M} = \BMOD$ and assignment \ASSI\ be given.
For $\varphi\in\FANG$, the proposition \WTF{\varphi}{\MO{M}}{\ASSI} is
defined as follows.
\begin{enumerate}
\item If $\varphi$ is $Pt_1 \dots t_n$ for $P\in\SIGN$ and terms
$t_1 \dots t_n$ then:\[\WTF{\varphi}{\MO{M}}{\ASSI} = \{w\in\WORLDS\MID
\langle \ASSI(t_1)\dots\ASSI(t_n)\rangle\in t(w,P)\}.\]
\item If $\varphi$ is the negation $\neg\theta$ then
$\WTF{\varphi}{\MO{M}}{\ASSI} = \WORLDS \setminus \WTF{\theta}{\MO{M}}{\ASSI}$.
\item If $\varphi$ is the conjunction $(\theta\wedge\psi)$ then
$\WTF{\varphi}{\MO{M}}{\ASSI} = \WTF{\theta}{\MO{M}}{\ASSI} \cap \WTF{\psi}{\MO{M}}{\ASSI}$.
\item If $\varphi$ is the existential $\exists x\psi$ then
$\WTF{\varphi}{\MO{M}}{\ASSI}$ is the set of $w\in\WORLDS$ such that
$w \in\WTF{\psi}{\MO{M}}{\ASSI'}$ for some $x$ variant $\ASSI'$ of \ASSI.
\item\label{defSat2a}
If $\varphi$ has the form $(\theta\succeq_X\psi)$ for $X\in\UIND$, then
$\WTF{\varphi}{\MO{M}}{\ASSI} = \emptyset$ if either
$\WTF{\theta}{\MO{M}}{\ASSI} = \emptyset$ or $\WTF{\psi}{\MO{M}}{\ASSI} = \emptyset$. Otherwise:
\[\WTF{\varphi}{\MO{M}}{\ASSI} = \{w\in\WORLDS\mid \UU_X(s(w,\WTF{\theta}{\MO{M}}{\ASSI})) \ge
\UU_X(s(w,\WTF{\psi}{\MO{M}}{\ASSI}))\}.\]
\end{enumerate}
\end{defn}
Thus, relative to \MO{M} and $d$, the formula $(\theta\succeq_X\psi)$
expresses the null proposition if evaluating it requires
that $s$ choose a world from $\emptyset$. (Preference makes a covert existential
claim in the present theory, namely, that there is something to choose between.)
Otherwise $w\in\WORLDS$ belongs to the proposition expressed by
$(\theta\succeq_X\psi)$  just in case the world chosen by $s$ to represent
\WTF{\theta}{\MO{M}}{\ASSI} has greater $X$-utility than the world
chosen by $s$ to represent \WTF{\psi}{\MO{M}}{\ASSI} --- where $s$'s choices
depend on the current situation $w$. Informally, we think of $s$ as choosing the most
similar world to $w$ among those available in the proposition at issue.

We extract the assignment-invariant core of a formula's proposition in the standard way.

\begin{defn}\label{meaning2}
Let $\varphi\in\FANG$ and model $\MO{M} = \BMOD$ be given. We write
\WF{\varphi}{\MO{M}} for the intersection of
\WTF{\varphi}{\MO{M}}{\ASSI} over all assignments \ASSI.
\end{defn}

\noindent
It follows that for closed $\varphi\in\FANG$ (no free variables),
$\WF{\varphi}{\MO{M}} = \WTF{\varphi}{\MO{M}}{\ASSI}$ for any assignment \ASSI.

\subsection{Global modality}

We can express global modality in the sense of
\citet[\S 2.1]{Blackburn} in the following way.
Choose any $X\in\UIND$, and for $\varphi\in\FANG$ define:
\begin{dispar}\label{boxdef}
\[
\BB\varphi \overset{\normalfont{\text{def}}}{=}
\neg(\neg\varphi \succeq_X\neg\varphi) \quad \text{and} \quad
\BD\varphi \overset{\normalfont{\text{def}}}{=}
(\varphi \succeq_X\varphi).
\]
\end{dispar}
\noindent
Then unwinding clause \ref{meaningDef}\ref{defSat2a} of our semantic definition
yields:
\begin{prop}\label{globalX}
For all $\varphi\in\FANG$, models $\MO{M} = \BMOD$, and assignments \ASSI:
\begin{enumerate}
\item\label{globalaX}
$\WTF{\BB\varphi}{\MO{M}}{\ASSI}\neq\emptyset$ iff $\WTF{\BB\varphi}{\MO{M}}{\ASSI} =\WORLDS$ iff $\WTF{\varphi}{\MO{M}}{\ASSI} = \WORLDS$.
\item\label{globalbX}
$\WTF{\BD\varphi}{\MO{M}}{\ASSI}\neq\emptyset$ iff $\WTF{\BD\varphi}{\MO{M}}{\ASSI} =\WORLDS$ iff $\WTF{\varphi}{\MO{M}}{\ASSI} \neq\emptyset$.
\end{enumerate}
\end{prop}
As an immediate and more perspicuous corollary, we have:
\begin{cor}\label{global}
For all closed $\varphi\in\FANG$, models $\MO{M} = \BMOD$:
\begin{enumerate}
\item\label{globala}
$\WF{\BB\varphi}{\MO{M}}\neq\emptyset$ iff $\WF{\BB\varphi}{\MO{M}} =\WORLDS$ iff $\WF{\varphi}{\MO{M}} = \WORLDS$.
\item\label{globalb}
$\WF{\BD\varphi}{\MO{M}}\neq\emptyset$ iff $\WF{\BD\varphi}{\MO{M}} =\WORLDS$ iff $\WF{\varphi}{\MO{M}} \neq\emptyset$.
\end{enumerate}
\end{cor}

\section{Basic properties of quantified preference logic}

\subsection{Expressive power of modal formulas}\label{effSec}

It is worth verifying that our modal vocabulary allows additional
propositions to be expressed.

\begin{defn}\label{modDep}
\begin{enumerate}
\item
The \textit{modal depth} of formulas is defined
inductively. First-order (non-modal) formulas have modal depth zero.
If $\varphi,\psi\in\SIGN$ have respective modal depths $m,n$ then
$\varphi\succeq_X\psi$ has modal depth $1 + \max\{m,n\}$.
\item
We say that a model \MO{M} \textit{has a modal hierarchy} just in case
there are closed formulas $\varphi_0, \varphi_1 \dots$ such that
for all $n \ge 0$:
\begin{enumerate}
\item\label{mhi1} $\varphi_n$ has modal depth $n$;
\item\label{mhi2} for all closed $\psi\in\LANG$ of modal depth $n$
or less, $\WF{\varphi_{n+1}}{\MO{M}} \neq \WF{\psi}{\MO{M}}$.
\end{enumerate}
\end{enumerate}
\end{defn}

\begin{defn}
Let $\MO{N} = \BNOD$ be the first three components of a model, missing just the
utility and selection functions, $u, s$. Notice that \BNOD\ assigns a proposition $\WF{\psi}{\MO{N}}\subseteq\WORLDS$
to each non-modal $\psi\in\LANG$. We call \MO{N} a
\textit{normal core} just in case
$D$ is countable,
\WORLDS\ is countably infinite, and there is
non-modal, closed $\psi\in\LANG$ with $\emptyset\neq\WF{\psi}{\MO{N}}\neq\WORLDS$.
\end{defn}

Now fix a countable signature \ASIGN.
The following proposition reveals the near ubiquity of modal hierarchies.

\begin{prop}\label{eff1}
Let $\MO{N} = \BNOD$ be a normal core.
Then there is a utility function $u:\UIND\times\WORLDS\to\Re$
and a selection function $s:\WORLDS\times\{A\subseteq\WORLDS\MID A\neq\emptyset\}\to\WORLDS$
such that the model $\MO{M} = \BMOD$ has a modal hierarchy.
\end{prop}

\noindent
\textsc{Proof}:
%
Choose utility index $X\in\UIND$,
let $\MO{N} = \BNOD$ be a normal core, and fix closed, non-modal $\psi\in\LANG$ with $\emptyset\neq\WF{\psi}{\MO{N}}\neq\WORLDS$. By replacing $\psi$ with its negation if necessary,
we can ensure that $\WF{\psi}{\MO{N}}$ has at least two elements.
Let $\varphi_0$ be $\psi$ and let $\varphi_{n+1}$ be $(\TAUT\prec_X\varphi_n)$.
Observe that for all $n\in\NATN$, $\varphi_n$ has modal depth $n$.
We will define $s$ and $u$ in such a way that $\varphi_0, \varphi_1 \dots$ is a modal hierarchy for $\MO{M}=\BMOD$.

Let $\{w_0,w_1,\dots\}$ enumerate \WORLDS. Since $\WF{\psi}{\MO{N}} = \WF{\varphi_0}{\MO{N}}$
has at least two elements, we may assume without loss of
generality that $\{w_0,w_1\}\subseteq\WF{\varphi_0}{\MO{N}}$. Let $u$ be any utility function
that meets the conditions:
\begin{dispar}\label{uthing}
$u_X(w_0)=0$ and for all $i>0$, $u_X(w_i)=1$.
\end{dispar}
It remains to specify the selection function $s$, and to show that it generates a modal
hierarchy. This is achieved by inductively defining a sequence of
``partial selection'' functions $s_n$, $n\in\NATN$. At stage $n$, the partial selector
$s_n$ defines a
partial model $\MO{M}_n = \PARMOD$ which yields a proposition
$\WTF{\chi}{\MO{M}_n}{\ASSI}$ for each assignment \ASSI, and
each $\chi\in\LANG$ of modal depth $n$ or below.
It will be easy to see that for each such $\chi$ and \ASSI,
$\WTF{\chi}{\MO{M}_n}{\ASSI} = \WTF{\chi}{\MO{M}}{\ASSI}$ where
$\MO{M} = \BMOD$ with $\bigcup_n s_n \subset s$. Let $\PFAM_n$ denote the family of
nonempty propositions
expressed by formulas of modal depth $n$ or below
with arbitrary assignments of members of $D$ to their free variables.
It is easy to verify that $\PFAM_n$ is countable.
At stage $n = 0$, we let $s_0 = \emptyset$.

For stage $n+1$, we will define $s_{n+1}$ so that:
\begin{enumerate}
\item\label{fcc} $s_{n+1}$
is defined for every pair $(w,X)$ where $w\in\WORLDS$ and $X\in \PFAM_n$; hence,
for every assignment \ASSI\ and $\chi\in\LANG$ of modal depth $n$ or below,
$\WTF{\chi}{\MO{M}_n}{\ASSI}$ is well defined.
\item\label{fca} $\WF{\varphi_{n+1}}{\MO{M}_{n+1}}\not\in \PFAM_{n}$ hence
$\WF{\varphi_{n+1}}{\MO{M}}\not\in \PFAM_{n}$;
\end{enumerate}

\noindent
Moreover, at every stage $n$, it will be the case that
$\{w_0,w_1\}\subseteq \WF{\varphi_n}{\MO{M}_n}$. In particular,
$\{w_0,w_1\}\subseteq \WF{\varphi_0}{\MO{M}_0} = \WF{\psi}{\MO{N}}$
follows from our choice of $\psi$.

Now we complete stage $n+1$.
For all $w\in\WORLDS$, set $s_{n+1}(w,\WORLDS) = w_0$ (hence we
always draw $w_0$ from the proposition expressed by \TAUT).
For all $w\in\WORLDS$ and all
$C\in\PFAM_n - \{\WF{\varphi_n}{\MO{M}_n},\WORLDS\}$, choose $s_{n+1}(w,C)$ to
be an arbitrary member of $C$.
For the remainder of $s_{n+1}$,
choose
$A\subseteq\WORLDS - \{w_0,w_1\}$ such that $A\not\in
\{B-\{w_0,w_1\}\MID B\in \PFAM_{n}\}$. Such an $A$ exists
because $\PFAM_n$ is countable.
For all $w\in\WORLDS$, we define:
\begin{equation*}
s_{n+1}(w,\WF{\varphi_n}{\MO{M}_n}) =
\begin{cases}
w_1  & \text{if } w\in A\cup \{w_0,w_1\}
\\
w_0 & \text{otherwise.}
\end{cases}
\end{equation*}

\noindent
It follows immediately from \ref{uthing} that
$\WF{\varphi_{n+1}}{\MO{M}_{n+1}}= A \cup\{w_0,w_1\}\not\in\PFAM_n$.
\QED

A natural question about Proposition \ref{eff1} is whether modal
hierarchies still appear when models satisfy various
\textit{frame properties}. To illustrate, model
$\MO{M} = \BMOD$ is called ``reflexive'' just in case for all
$w\in\WORLDS$ and $A\subseteq\WORLDS$, if $w\in A$ then $s(w,A) = w$.
Reflexivity embodies the idea that the actual world is closer to
home than any other world. Several frame properties are examined
in \cite{PBOR}, and also below. In the case of reflexivity, the
foregoing proof can be adjusted to show that any normal core
can be extended to a reflexive model with modal hierarchy.
We leave unexplored the larger project of characterizing the
frame properties that allow modal hierarchies, or identifying
natural properties that do not.

\subsection{Undecidability of satisfaction}

Suppose that the signature \ASIGN\ contains
two unary predicates $P,Q\in\SIGN$. Then it follows from the argument
in \cite{kripke} that:

\begin{prop}\label{und1}
The satisfiable subset of \FANG\ is not decidable.
\end{prop}

\noindent Kripke's argument hinges on a mapping from
first-order sentences with just the binary relation symbol
$R$ to modal sentences that replace $Rxy$ with $\BD(Px \wedge Qy)$.
On the other hand, the validities are axiomatizable:

\begin{prop}\label{ce-propA}
If the signature is effectively enumerable then so is the set of
valid formulas in quantified preference logic.
\end{prop}

\noindent
This fact follows from Proposition \ref{ce-prop}, below.

\subsection{Size of models}

Suppose that the signature contains a binary predicate $G$. Then the
upward L\"{o}wenheim-Skolem property fails to apply to quantified
preference logic. Indeed:

\begin{prop}\label{secondTry}
There is $\varphi\in\FANG$ such that:
\begin{enumerate}
\item\label{secondTrya} Some model \BMOD\ with $D$ countable
satisfies $\varphi$.
\item\label{secondTryb} No model \BMOD\ with $D$ uncountable
satisfies $\varphi$.
\end{enumerate}
\end{prop}

\noindent\textsc{Proof}: Basically, $\varphi$ says that $\prec$ is a lexicographical
order on $D\times D$; such an order cannot be
embedded in $\langle \Re,<\rangle$ if $D$ is uncountable.
For typographical simplicity, we choose $X\in\UIND$, and write $\prec$ in place of $\prec_X$.

Specifically,
we take $\varphi$ to be the conjunction of the following formulas.
\begin{dispar}\label{ff}
\begin{enumerate}
\item\label{ffe} $\forall x \forall y (x \neq y \rightarrow ((Gxx \prec Gyy) \vee (Gyy \prec Gxx))$
\item\label{ffd} $\forall x_1y_1x_2y_2((Gx_1y_1 \prec Gx_2y_2)\leftrightarrow ((Gx_1x_1 \prec Gx_2x_2)
\vee ((x_1 = x_2) \wedge (Gy_1y_1\prec Gy_2y_2))))$
\end{enumerate}
\end{dispar}
Let a model $\MO{M} = \BMOD$ and $w_0\in\WORLDS$ be given with $w_0\in\WF{\varphi}{\MO{M}}$.
We define:
\[X=\{u(s(w_0,\WTF{Gxx}{\MO{M}}{\ASSI(a/x)}))\MID a\in D\}.\]
Then \ref{ff}\ref{ffe}
implies that $X$ (a set of reals) has the same cardinality as $D$.
Define:
\[Y=\{u(s(w_0,\WTF{Gxy}{\MO{M}}{\ASSI(a/x,b/y)}))\MID a,b\in D\}.\]
Then \ref{ff}\ref{ffd} implies that $\langle Y, <\rangle$ is isomorphic to
the lexicographic ordering of $X\times X$.

We leave to the reader
the verification that $\varphi$ is satisfiable in
a model with countable domain.
On the other hand, suppose that the domain is
uncountable, whence $X$ is uncountable. Then the existence of an isomorphism between $\langle Y, <\rangle$
and the lexicographic ordering of $X\times X$ contradicts the separability of the real line. \QED

\subsection{Preorder models}\label{preorder-sec}
We can recover the upward L\"{o}wenheim-Skolem property by introducing
a more general way to compare the value of worlds.  Recall that a (\textit{total}) \textit{preorder} is
transitive, connected, and reflexive over its domain. Given a signature \ASIGN,
we achieve more generality by replacing $u$ in a model
\BMOD\ with a map \PRO\ from \UIND\
to the set of preorders over \WORLDS.
[We write $\PRO_X$ for $\PRO(X)$, $X\in\UIND$.] In such a
model \BOOD, we evaluate $(\theta\succeq_X\psi)$ according to the following rule,
in place of \ref{meaningDef}\ref{defSat2a}.

\begin{quote}
\ref{meaningDef}\ref{defSat2a}$'$
If $\varphi$ has the form $(\theta\succeq_X\psi)$ for $X\in\UIND$, then
$\WTF{\varphi}{\MO{M}}{\ASSI} = \emptyset$ if either
$\WTF{\theta}{\MO{M}}{\ASSI} = \emptyset$ or $\WTF{\psi}{\MO{M}}{\ASSI} = \emptyset$.
Otherwise:
\[\WTF{\varphi}{\MO{M}}{\ASSI} = \{w\in\WORLDS\mid
s(w,\WTF{\theta}{\MO{M}}{\ASSI})\,\PRO_X\,
s(w,\WTF{\psi}{\MO{M}}{\ASSI})\}.\]
\end{quote}

\noindent
In what follows, we'll call the semantics based on \ref{meaningDef}\ref{defSat2a}$'$
\textit{preorder logic}. The original semantics, based on \ref{meaningDef}\ref{defSat2a},
will be called \textit{utility logic}.
It is easy to see that utility logic is a special case
of preorder logic (since assigning utilities to worlds preorders them).
Also, it is straightforward to show that
the formula $\varphi$ in the proof of Proposition
\ref{secondTry} is satisfied in a preorder model with uncountable domain $D$.
Indeed, the following L\"{o}wenheim-Skolem Theorem holds for preorder models.
\begin{prop}\label{lst-prop}
Let \BOOD\ be a preorder model for a countable signature.
\begin{enumerate}
\item If $\WORLDS$ is infinite, then for every infinite cardinal $\kappa$ there is a preorder model $\MO{M'}=\langle D',\WORLDS',t',\PRO',s'\rangle$ such that $\mathsf{card}(\WORLDS')=\kappa$ and for every sentence $\varphi$,
\[\MO{M}\models\varphi \text{ if and only if } \MO{M'}\models\varphi.\]
\item If $D$ is infinite, then for every infinite cardinal $\kappa$ there is a preorder model $\MO{M'}=\langle D',\WORLDS',t',\PRO',s'\rangle$ such that $\mathsf{card}(D')=\kappa$ and for every sentence $\varphi$,
\[\MO{M}\models\varphi \text{ if and only if } \MO{M'}\models\varphi.\]
\end{enumerate}
\end{prop}

Despite the greater generality of preorder logic, and the contrast between Propositions \ref{lst-prop} and \ref{secondTry},
the distinction between utility and preorder models is not discernible by formulas. Indeed:
\begin{prop}\label{uvispv-prop}
A formula $\theta$ is valid in the class of utility models if and only if it is valid in the class of preorder models.
\end{prop}

\noindent
Finally, the next proposition shows that the set of formulas which are valid in preorder models (and hence utility models,
by the preceding proposition) is axiomatizable. We assume that the signature is effectively enumerable.
\begin{prop}\label{ce-prop}
The set of formulas which are valid in preorder models is effectively enumerable.
\end{prop}

\noindent
Proofs of Propositions \ref{lst-prop}, \ref{uvispv-prop}, and \ref{ce-prop} are given in the Appendix.

\subsection{Generalized preference logic}

Reliance on utilities to express preference has a long history in economics
e.g., \citet{vnm}. Proposition \ref{uvispv-prop} provides some justification
for this approach inasmuch as the more general preorder logic does not change
the class of validities, compared to utility logic. Indeed, validity is preserved in
utility logic even if the range of $u$ is limited to the rationals (since every
countable preorder is isomorphic to the natural ordering of a subset of rationals).

A yet more general logic is presented in \cite{PBOR}.
We call a quadruple \GOOD\ a \textit{generalized model} for signature \ASIGN\
if $D$, \WORLDS, and $t$ are as before, and $\GRO$ is a mapping
from $\WORLDS\times\UIND$ to the set of preorders over nonempty
propositions (nonempty subsets of \WORLDS). The selection function
$s$ no longer appears since propositions are now compared directly
rather than via representative worlds chosen by $s$.
In this kind of model
we evaluate $(\theta\succeq_X\psi)$ according to the following variant
of \ref{meaningDef}\ref{defSat2a}.

\begin{quote}
\ref{meaningDef}\ref{defSat2a}$''$
If $\varphi$ has the form $(\theta\succeq_X\psi)$ for $X\in\UIND$, then
$\WTF{\varphi}{\MO{M}}{\ASSI} = \emptyset$ if either
$\WTF{\theta}{\MO{M}}{\ASSI} = \emptyset$ or $\WTF{\psi}{\MO{M}}{\ASSI} = \emptyset$. Otherwise:
\[\WTF{\varphi}{\MO{M}}{\ASSI} = \{w\in\WORLDS\mid
\WTF{\theta}{\MO{M}}{\ASSI} \text{ comes no earlier than }
\WTF{\psi}{\MO{M}}{\ASSI}\text{ in }
\GRO(w,X)\}.\]
\end{quote}

\noindent
In \cite{PBOR} we analyze the relation between generalized models
and utility models in the sentential context. For quantified preference
logic, matters are less clear and we leave the following question open.

\begin{opqu}\label{gculog}
Are the set of formulas valid in utility models the same as the
formulas valid in generalized models?
\end{opqu}
An affirmative answer would provide further evidence for the sufficiency
of the utility approach to preference. Of course, validity might be
preserved between logics when all their models are considered but
break down if attention is limited to certain subsets. The next
section illustrates subsets of models that are defined by
natural properties. For the remainder
of the discussion, only utility models (introduced in Section
\ref{semSec}) are at issue.

\section{Subclasses of utility models}

\subsection{Metricity}

Many interesting properties of a model \BMOD\ can formulated just in
terms of \WORLDS\ and $s$ (the model's ``frame''). For example,
\cite{PBOR} consider the following way to express the idea that
$s$ chooses ``the nearest world.''

\begin{defn}\label{metric1}
A model \BMOD\ is \textit{metric}
just in case there is a metric $d\colon \WORLDS\times\WORLDS
\to \Re$ such that for all $w\in\WORLDS$ and $\emptyset\neq
A\subseteq\WORLDS$, $s(w,A)$ is the unique $d$-closest member
of $A$ to $w$.
\end{defn}
\noindent
Note that a model is metric only if $d$-closest worlds exist
(there are no chains of worlds ever $d$-closer to a given world).
It is easy to see that in a metric model the set of worlds is
countable. There are several properties of models that
are implied by metricity. Here we focus on:

\begin{defn}\label{transdef}
A model \BMOD\ is \textit{\TRA} just in case for all
$A,B,C\subseteq\WORLDS$ with $A,B\neq\emptyset$, and
$w_0\in\WORLDS$, if $s(w_0,A\cup B)\in A$ and $s(w_0,B\cup C)\in B$ then
$s(w_0,A\cup C) = s(w_0,A\cup B)$.
\end{defn}
Exploiting our quantificational apparatus, we can write a formula
that is true in all \TRA\ models but not valid. We assume that
the signature includes the predicate $P$.
For notational ease, we suppress the $X$ on $\approx_X$.

\begin{prop}\label{traProp}
Let $\varphi$ be the conjunction of the following formulas.
\begin{enumerate}
\item\label{traPropa}
$\forall xy(x\neq y \rightarrow (Px \not\approx Py))$
\item\label{traPropb}
$\forall xyz\,((x\neq y \wedge y \neq z \wedge x \neq z)\rightarrow ((((Px\vee Py) \approx Px) \wedge ((Py\vee Pz)\approx Py))\rightarrow
(Px\vee Pz)\approx Px))$
\end{enumerate}
Then $\varphi$ is invalid but valid in the class of \TRA\ models.
\end{prop}
\noindent
The proposition can be viewed as expressing the transitivity of revealed preference, e.g.,
$(Px\vee Py) \approx Px$
says that $Px$ is chosen from the mutually exclusive options $Px, Py$.

\textsc{Proof}: Let model $\MO{M} = \BMOD$, $w_0\in\WORLDS$ and assignment \ASSI\ be given.
Let $\WTF{Px}{\MO{M}}{\ASSI} = A$,
$\WTF{Py}{\MO{M}}{\ASSI} = B$ and $\WTF{Pz}{\MO{M}}{\ASSI} = C$. If any of
$\ASSI(x), \ASSI(y), \ASSI(z)$ are identical
or either $A$ or $B$ are empty then
we are done. Otherwise, in the presence of \ref{traProp}\ref{traPropa}, $(Px\vee Py) \approx Px$
and $(Py\vee Pz) \approx Py$
imply respectively that $s(w_0,A\cup B) \in A$ and $s(w_0,B\cup C) \in B$.
So \TRAL\ implies $s(w_0,A\cup C) = s(w_0,A\cup B)$
which entails $w_0\in\WTF{(Px\vee Pz)\approx (Px \vee Py)}{\MO{M}}{\ASSI}$. So the proposition
follows by the transitivity of $\approx$ from $w_0\in\WTF{(Px\vee Py)\approx Px}{\MO{M}}{\ASSI}$. \QED

\subsection{Beyond the frame}

Rational agents might not be able to discriminate between isomorphic worlds.
To formulate this idea,
fix a signature \ASIGN, and let
model $\MO{M} = \BMOD$ be given. We say that $v,w \in \WORLDS$ are
\textit{isomorphic} ($v \simeq w$) just in case there is a permutation $h$ of $D$
such that for all $Q\in\SIGN$,
$h$ (applied component-wise) maps $t(v,Q)$ onto $t(w,Q)$.

\begin{defn}\label{def1}
Model \BMOD\
is \textit{\INVU} just in case
for all isomorphic $v,w\in\WORLDS$,
$u_X(v) = u_X(w)$ for all $X\in\UIND$.
\end{defn}
This is not a frame property because all components of the model are involved in
its formulation. Validity in the \INVU\ models doesn't
imply validity in the strict sense.
Indeed, we have:

\begin{prop}\label{provInv2}
Let signature \ASIGN\ be given with \SIGN\ finite, and
distinct $X,Y\in\UIND$. Then
there is invalid $\varphi\in\FANG$ that is valid in the class of \INVU\ models.
\end{prop}

\textsc{Proof}: There is $\chi\in\FANG$
such that for all models $\MO{M} = \BMOD$,
$\WF{\chi}{\MO{M}} = \WORLDS$ iff $\mid D\mid\  = 2$. Hence,
by the finitude of \SIGN\ and the presence of identity,
there is closed, satisfiable
$\psi\in\FANG$ such that for all models \MO{M}, if
$w_1, w_2 \in \WF{\psi}{\MO{M}}$ then $w_1\simeq w_2$.
Let the promised $\varphi$ be:
\[(\psi \wedge (\psi\succ_X\TAUT) \wedge (\psi\succ_Y\TAUT)) \rightarrow
((\psi \wedge(\psi\succ_X\TAUT)) \thickapprox_X (\psi \wedge(\psi\succ_Y\TAUT))).\]
We indicate why $\varphi$ is invalid. The antecedent of
$\varphi$ is easily seen to be satisfiable, and a $\psi$-world satisfying
$\psi \wedge(\psi\succ_X\TAUT)$
need not be the same world that satisfies $\psi \wedge(\psi\succ_Y\TAUT))$;
and $u_X$ may be chosen to be injective.

On the other hand, suppose that model $\MO{M} = \BMOD$ is \INVU\ and
let $w_0 \in \WORLDS$. Suppose that the antecedent of $\varphi$
is satisfiable in \MO{M} (otherwise, we are done). Then $\WF{(\psi \wedge(\psi\succ_X\TAUT)}{\MO{M}}
\neq\emptyset$ and $\WF{(\psi \wedge(\psi\succ_Y\TAUT))}{\MO{M}}\neq\emptyset$. So,
let $w_1 = s(w_0,\WF{(\psi \wedge(\psi\succ_X\TAUT))}{\MO{M}})$ and
$w_2 = s(w_0,\WF{\psi \wedge(\psi\succ_Y\TAUT)))}{\MO{M}})$.
Then each of $w_1, w_2$ satisfies $\psi$ so $w_1\simeq w_2$.
Hence $u_X(w_1) = u_X(w_2)$ by \INVUL. \QED

\section{Anonymity}

Our final topic concerns the manner in which utilities are associated
with formulas. First, a condition is exhibited that makes
the utility of a conjunction depend on just the utilities of each conjunct
separately. For example, according to this condition
the vocabulary appearing in a conjunct is not
permitted to influence the utility of the conjunction; rather,
the conjunct contributes its utility ``anonymously.''
A second condition is then introduced that
entails a similar kind of anonymity for the contribution of utility
indexes $1$ and $2$ to the aggregated utility $\{1,2\}$.
The material in this section is inspired by the discussion in
\citet[\S 7.2]{KLST}.

\subsection{Decomposing the utility of conjunctions}
Let a signature \ASIGN\ be given with predicate $P\in\SIGN$.
Conjunctive anonymity with respect to $P$
is expressed by the following formula.
(To lighten notation, we suppress $X\in\UIND$ in subscripts.)

\begin{dispar}\label{dc4}
$\varphi\ \NICEDEF\  \forall xy((Px \approx Py) \rightarrow \forall z((Px \wedge Pz) \approx (Py \wedge Pz)))$
\end{dispar}

\noindent
The next proposition gives the sense in which $\varphi$ causes the utility of $Px \wedge Py$ to be a function
($F$) of the utilities of $Px$ and $Py$.

\begin{prop}\label{dc6}
Let model $\MO{M}=\BMOD$ be given with
$w_0\in\WF{\varphi}{\MO{M}}$. Then there is a function
$F:\Re^2\to\Re$ such that for all assignments \ASSI\ with
$\WTF{Px \wedge Py}{\MO{M}}{\ASSI}\neq\emptyset$,

\[
u(s(w_0,\WTF{Px \wedge Py}{\MO{M}}{\ASSI})) = F(\,u(s(w_0,\WTF{Px}{\MO{M}}{\ASSI})),\,u(s(w_0,\WTF{Py}{\MO{M}}{\ASSI}))\,).
\]
\end{prop}

\textsc{Proof}:
For numbers of the form $u(s(w_0,\WTF{Px}{\MO{M}}{\ASSI}))$ and
$u(s(w_0,\WTF{Py}{\MO{M}}{\ASSI}))$ define:
\begin{dispar}\label{cd22}$
F(\,u(s(w_0,\WTF{Px}{\MO{M}}{\ASSI})),\,u(s(w_0,\WTF{Py}{\MO{M}}{\ASSI}))\,)\NICEDEF
u(s(w_0,\WTF{Px \wedge Py}{\MO{M}}{\ASSI}))$.
\end{dispar}
For all other numbers $r_1, r_2$, $F(r_1,r_2)$ is defined arbitrarily.
We must show that $F$ is a function. For this purpose, let
variable $q$ be given, and suppose that
\begin{dispar}\label{cd10}
$u(s(w_0,\WTF{Px}{\MO{M}}{\ASSI})) = u(s(w_0,\WTF{Pq}{\MO{M}}{\ASSI}))$.
\end{dispar}
To finish the proof it suffices to show that
\begin{dispar}\label{cd11}
$u(s(w_0,\WTF{Px \wedge Py}{\MO{M}}{\ASSI})) = u(s(w_0,\WTF{Pq \wedge Py}{\MO{M}}{\ASSI}))$,
\end{dispar}
the second argument of $F$ being treated in the same way.
It follows immediately from \ref{cd10} that
$
w_0\in\WTF{(Px \approx Pq)}{\MO{M}}{\ASSI}$,
hence by \ref{dc4}
\[
w_0\in\WTF{((Px\wedge Py) \approx (Pq\wedge Py))}{\MO{M}}{\ASSI},
\]
which implies \ref{cd11}. \QED

Observe that $\varphi$ and Proposition \ref{dc6} can be
formulated with disjunction in place of conjunction --- or with
many other formulas. The proof proceeds in the same way.

\subsection{Decomposing a complex utility index}

Suppose for this section that the signature \ASIGN\ contains
unary $P \in \SIGN$ along with
$\{1\}, \{2\}, \{1,2\}\in\UIND$. Define:

\begin{dispar}\label{di1}
$\varphi\ \NICEDEF\  \forall xy(\,((Px \approx_1 Py) \wedge (Px \approx_2 Py))\rightarrow (Px \approx_{1,2} Py)\,)$.
\end{dispar}

\noindent
Then $\varphi$ implies that the contributions of $1$ and $2$ to the complex utility
index $\{1,2\}$ can be separated then brought back together via a binary mapping on
$\Re$. Specifically:

\begin{prop}\label{di2}
Let model $\MO{M}=\BMOD$ be given with $w_0\in\WF{\varphi}{\MO{M}}$.
Then there is a function
$F:\Re^2\to\Re$ such that for all assignments \ASSI:
\[
u_{1,2}(s(w_0,\WTF{Px}{\MO{M}}{\ASSI})) = F(\,u_{1}(s(w_0,\WTF{Px}{\MO{M}}{\ASSI})), \,u_{2}(s(w_0,\WTF{Px}{\MO{M}}{\ASSI}))\,).
\]
\end{prop}

\textsc{Proof}:
Call a pair $(p,q)\in\Re^2$ \textit{critical} just in case there is an assignment
\ASSI\ such that
\begin{dispar}\label{di3}
\begin{enumerate}
\item\label{di3a} $p = u_{1}(s(w_0,\WTF{Px}{\MO{M}}{\ASSI}))$
\item\label{di3b} $q = u_{2}(s(w_0,\WTF{Px}{\MO{M}}{\ASSI}))$.
\end{enumerate}
\end{dispar}
Let $F:\Re^2\to\Re$ be such that for any critical pair $(p,q)$ as in\ref{di3},
$F(p,q) = u_{1,2}(s(w_0,\WTF{Px}{\MO{M}}{\ASSI}))$. The behavior of $F$ on
noncritical pairs is arbitrary. Suppose that for some assignment $\ASSI'$:
\begin{dispar}\label{di4}
\begin{enumerate}
\item\label{di3ax} $p = u_{1}(s(w_0,\WTF{Px}{\MO{M}}{\ASSI'}))$
\item\label{di3bx} $q = u_{2}(s(w_0,\WTF{Px}{\MO{M}}{\ASSI'}))$.
\end{enumerate}
\end{dispar}
To verify that $F$ is a function, thereby completing the proof, we must
show that
\begin{dispar}\label{di5}
$u_{1,2}(s(w_0,\WTF{Px}{\MO{M}}{\ASSI})) = u_{1,2}(s(w_0,\WTF{Px}{\MO{M}}{\ASSI'}))$.
\end{dispar}
Let $y$ be a variable distinct from $x$, and let $\ASSI'' = \ASSI(\ASSI'(x)/y)$.
From \ref{di3} and \ref{di4} we infer: $w_0 \in \WTF{Px \approx_1 Py}{\MO{M}}{\ASSI''}$ and
$w_0 \in \WTF{Px \approx_2 Py}{\MO{M}}{\ASSI''}$. From \ref{di1} we then obtain
$w_0 \in \WTF{Px \approx_{1,2} Py}{\MO{M}}{\ASSI''}$ from which \ref{di5} is an immediate
consequence. \QED

\section*{Appendix}

We present proofs of Propositions \ref{lst-prop}, \ref{uvispv-prop}, and \ref{ce-prop} from Section \ref{preorder-sec}.
All three proofs elaborate a construction that appears in the demonstration of Theorem (55) in \cite{PBOR}.
Specifically, the earlier
construction can be adapted to show that there is an effective translation from sentences $\varphi\in\FANG$ to
formulas $\trans{\varphi}(x)$ of first-order logic, and a map from preorder models $\MO{M}=\BOOD$ to  relational structures \RELF{M} such that
\begin{dispar}\label{translation}
$w\in\varphi[\MO{M}]\ \ \mbox{iff}\ \ \RELF{M}\models\trans{\varphi}[w]$.
\end{dispar}
Moreover, assuming that \ASIGN\ is recursive, there is a recursively axiomatizable first-order theory $T$ in the signature of \RELF{M}\ such that
\begin{dispar}\label{relfsatd}
for every preorder model \MO{M}, $\RELF{M}\models T$
\end{dispar}
and
\begin{dispar}\label{Dinv}
for every first-order structure $A$, if $A\models T$, then for some preorder model \MO{M}, $A=\RELF{M}.$
\end{dispar}
Proposition \ref{ce-prop} now follows from the completeness theorem for first-order logic, since \ref{translation}, \ref{relfsatd}, and \ref{Dinv} imply that $\varphi\in\FANG$ is valid in preorder logic if and only if $\forall x \trans{\varphi}(x)$ is a consequence of $T$. In like fashion, Proposition \ref{lst-prop} follows from the L\"{o}wenheim-Skolem Theorem for first-order logic. Proposition \ref{uvispv-prop} now follows immediately, since every countable preorder model is induced by a corresponding utility model, a consequence of the fact that the rational numbers are universal among countable linear orders. \QED

\renewcommand{\baselinestretch}{1.0}
\begin{small}
\bibliography{reasonbib}
\end{small}
\clearpage

\end{document}